\documentclass[12pt]{article}

\usepackage{amssymb}
\usepackage{epsfig,color}
\usepackage{pstricks,graphicx,epsfig,color,amssymb,amsmath,amscd}
\usepackage{cite}
\usepackage[]{graphicx}
\usepackage{makeidx}

\newcommand{\be}{\begin{eqnarray}}
\newcommand{\ee}{\end{eqnarray}}
\newcommand{\rar}{\rightarrow}

\usepackage[]{caption}
\renewcommand\rho{\varrho}

\date{}

\begin{document}

\begin{titlepage}
\title{Possible explanation of primordial {$^7$Li} deficit }
\author{E.V. Arbuzova$^{a,b}$, A.D. Dolgov$^{b,c}$}

\maketitle
\begin{center}
$^a${Dubna State University, Department of Higher Mathematics, \\Universitetskaya ul. 19, Dubna 141983, Russia}\\
$^b${Novosibirsk State University,  Department of Physics,\\Pirogova 2, Novosibirsk 630090, Russia}\\
$^c${Bogolyubov Laboratory of Theoretical Physics, Joint Institute for Nuclear Research,
Joliot-Curie st. 6, Dubna, Moscow region, 141980 Russia}

\end{center}

\begin{abstract}

A reduction mechanism  of the theoretically predicted excessive abundance of $^7$Li  via baryons evaporated 
by primordial black holes is suggested. It is shown that the fraction of $^7$Li with respect to the number density 
of baryons can be diminished down to the observed value
 via the process of $^7$Li transformation  by neutron capture. The created in this process $^8$Li or $^8$Be
 quickly decay into a pair of $^4$He nuclei.

\end{abstract}
\thispagestyle{empty}
\end{titlepage}

\section{Introduction}

Big Bang nucleosynthesis (BBN) is justly considered as one of the cornerstones of canonical cosmological model. In particular, it
predicts the observed mass fraction of $^4$He equal to 25\%, that cannot be explained otherwise. For a review see 
e.g. Ref.~\cite{ParticleDataGroup:2024cfk},  in particular  Fig.~24.1, presented below:

\begin{figure}[htbp]
		\begin{center}
		\includegraphics[scale=0.5,angle=0]{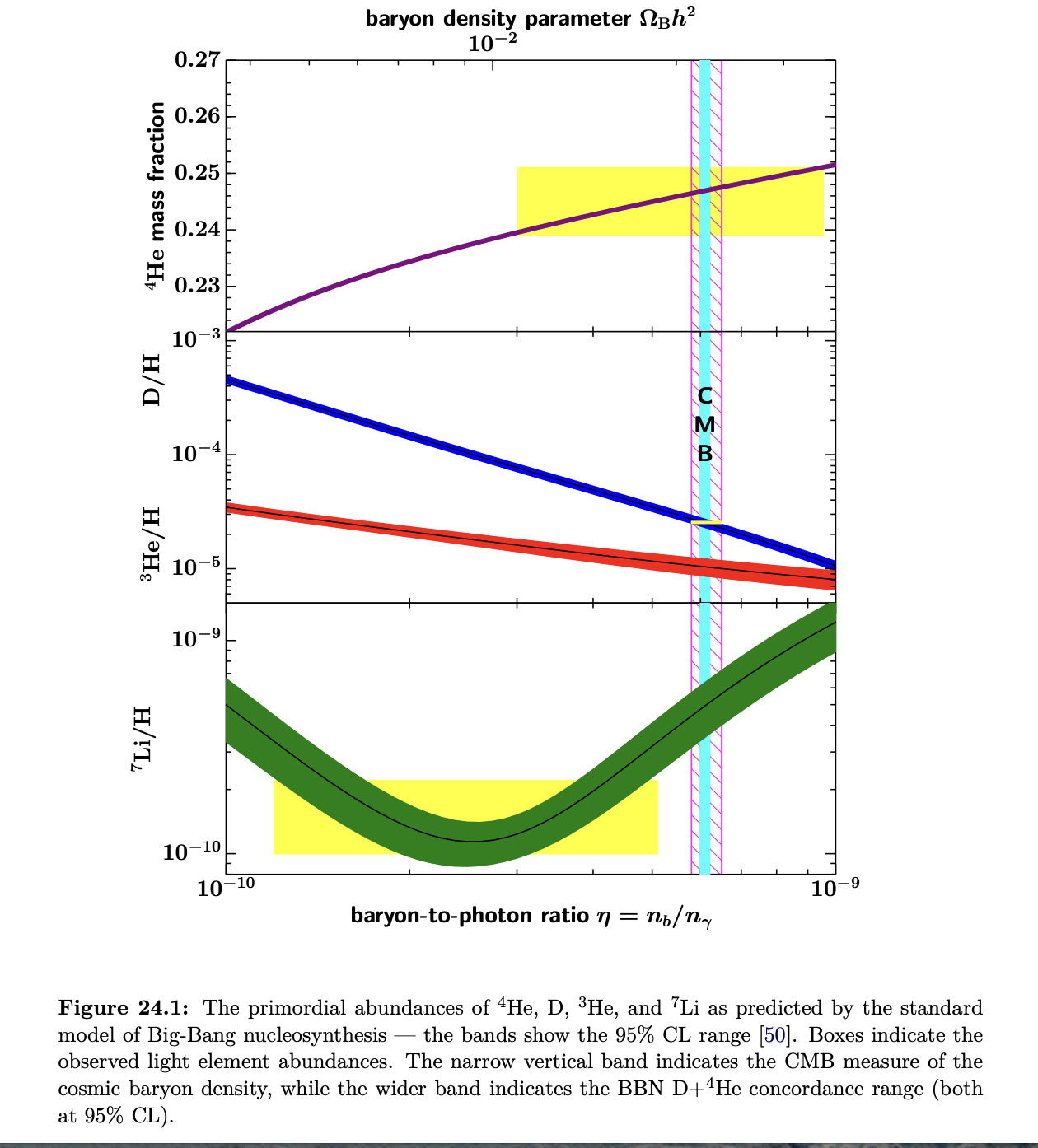} \hspace{2mm}
		\vspace{-3mm}
       \end{center}
	\caption{{{Fractions of light elements produced at BBN.
	}}}	\label{f-fig1}
\end{figure}
The observed abundances of $^4$He, deuterium, and $^3$He perfectly well agree with theoretical predictions, at the 
canonical value
of the baryon-to-photon ratio $\eta \approx  6\times 10^{-10} $ found from angular fluctuations of cosmic microwave 
background 
radiation, see e.g. Ref.~\cite{Planck:2018vyg}. 
However, the  theoretically predicted $^7$Li-fraction  at such $\eta$ exceeds the observed value, 
$^7$Li/H $\approx 1.6\times 10^{-10}$~\cite{ParticleDataGroup:2024cfk}, 
approximately by factor 3. 
The review of $^7$Li problem and possible, not yet satisfactory solutions, can be found in Ref.~\cite{Li-prob-rev}.

In the present work we suggest an alternative mechanism to explain the observed lithium-7 deficit via $^7$Li transformation into 
$^8$Li or $^8$Be by the capture of neutrons in the reactions: $\mathrm{^{7}Li} + n \rightarrow \mathrm{^{8}Li} + \gamma$ or 
$\mathrm{^{7}Li} + n \rightarrow \mathrm{^{8}Be} + \gamma$. The produced $^8$Li or $^8$Be
quickly decay  into a pair of $^4$He. We assume that the neutrons are created by the evaporation of primordial black holes (PBH) with 
properly adjusted parameters.

\section{A few words about (Schwarzschild) black hole evaporation 
\label{s-BH-evap}}

According to S. Hawking discovery~\cite{Hawking:1974rv} a black hole  (BH) with mass $M_{BH}$ has the temperature:
\be
T_{BH} = { M_{Pl}^2 \over 8\pi M_{BH}} = 1.05\cdot 10^3 \,{\rm GeV}/M_{10},  
\label{TBH}
\ee
where the Planck mass is $M_{Pl} = 2.176 \times 10^{-5}$ g $ = 1.22 \times 10^{19}$ GeV
and $M_{10} = M_{BH}/{( 10^{10} {\rm g}  )}$. All particles with masses below $T_{BH} $ are abundantly created in the
course of evaporation.

Since the BH temperature is inversely proportional to its mass and the black hole  area is proportional to the gravitational 
radius 
\be
r_g = 2 M_{BH} /M_{Pl}^2 \, ,
\label{r-g}
\ee
the energy flux of the emitted radiation is $J \sim T^4 r^2_g \sim ( M_{BH}^{-2})$, i.e.
$\dot M_{BH} \sim \left(-M^{-2}_{BH}\right)$.  Correspondingly the mass of BH, created at time moment $t^{(in)}$ evolves as
\be
M_{BH}(t) = M_{BH}^{(in)} \left[\frac{\tau_{BH} - t+t^{(in)}}{\tau_{BH}  } \right]^{1/3},
\label{M-of-t}
\ee
with the black hole life-time, as calculated in Ref.~\cite{Page:1976df}, equal to:
\be
\tau_{BH} \approx 3\times 10^3 N_{eff}^{-1} M_{BH}^3 M_{Pl}^{-4}  \equiv C\,\frac{M_{BH}^3}{M_{Pl}^4},
\label{tau-BH}
\ee 
{where $C \approx 30$}, and $N_{eff} \approx 100$ is the effective number of particle species with
masses smaller than the black hole temperature. In terms of seconds the BH life time is:
\be 
\tau_{BH} = 160\,M_{10}^3\,\,{\rm sec} .
\label{tau-10}
\ee

\section{Thermalisation}
\label{s-therm}

Baryons produced in the course of BH evaporation quickly loose their energy through interaction with the primeval plasma.
Indeed, the thermalisation/equilibration time can be estimated as:
\be
{t_{eq} = 1/\left( \sigma_{tot} v_{cosm} n_{cosm} \right),}
\label{t-eq}
\ee
where $n_{cosm}$ is the number density of particles in the primeval plasma, 
 $\sigma_{tot}$ is the total interaction cross-section of the baryons with the background matter,
and {{$v_{cosm} $ is  the relative velocity of the colliding particles in cosmological plasma.}}

{In thermal equilibrium, the particle number density  at temperature $T_{cosm}$ is:}
\be
n_{cosm} \approx 0.1 T_{cosm}^3 g_* , 
\label{n-plasma}
\ee
where $g_* \sim 10^2$ is the number of particle species. The numerical coefficient in Eq.~(\ref{n-plasma})
is equal to 0.09 for fermions and 0.12 for bosons, but these subtleties can here be neglected, 
 so  instead of 0.09 or 0.12 we use  0.1 in both cases.
Equilibrium is established if
the characteristic time is shorter than the Hubble expansion (cooling) time.
Indeed, according to the Friedman equation, the Hubble parameter, $H$, is expressed though the cosmological energy density 
(for $3D$ flat universe): 
\be
\frac{3 H^2 M_{Pl}^2}{8\pi} =  \rho_{cosm}. 
\label{rho-cosm}
\ee

The energy density of cosmological plasma at thermal equilibrium is related to cosmological temperature as
\be
\rho_{cosm} = \frac{\pi^2g_*}{30} T_{cosm}^4.
\label{rho-of-T}
\ee

Hence the Hubble cooling time as a function of cosmological temperature is equal to:
\be
\tau_{cosm} \equiv H^{-1} = \left(\frac{90}{8\pi^3 g_*}\right)^{1/2}\,\frac{M_{Pl}}{T^2_{cosm}} =
0.5 \,{\rm sec} \left({\rm MeV}/{T_{cosm}}\right)^2 .
\label{tau-cosm}
\ee

Baryons created by PBHs typically have energies of the order of the PBH temperature, i.e. about   
{ one} GeV in our cases.
However, the baryons are ejected { into} the low energy background plasma that by assumption has 
much larger { number and energy densities
than  the  densities of energetic baryons created by}  the evaporating primordial black holes. These energetic baryons would loose
energy during the characteristic thermalisation time 
\be
{\tau_{therm} = \left(\sigma_{tot} v_{evap} n_{cosm} \right)^{-1},}
\label{tau-therm}
\ee
where {{$n_{cosm}$} is given by Eq.~(\ref{n-plasma})} {and $v_{evap}$ is typical velocity of evaporating particles, which 
is initially close to the speed of light.}

For efficient thermalisation during the Hubble time $\tau_{cosm}$ (\ref{tau-cosm}) 
 the condition  { $\tau_{therm} < \tau_{cosm}$} should be fulfilled. { It is  satisfied for the cross-section
\be
\sigma_{tot} v_{evap} > 5\times 10^{-42} \left(\frac{\rm keV}{T_{cosm}}\right)\,{\rm cm}^2.
\label{thermalisation}
\ee }

This is surely true for the case we are interested in. In fact,
it is well known that thermal equilibrium in cosmology 
is established much faster than the Hubble expansion rate for any reasonable value of the interaction cross-section.



\section{Hot black holes \label{ss-hot-BH}}

Here we assume that the temperature of the considered PBHs is about 1 GeV. 
PBH with such temperature could efficiently create baryons and anti-baryons. 
The PBH mass in this case is about $10^{13}$ g and the lifetime 
$\tau_{BH} \sim1.6 \times 10^{2} (M_{BH}/ 10^{10} g)^3\, {\rm sec} \approx 1.6\times 10^{11}  $ sec, 
see Eq.~(\ref{tau-10}). 
According  Eq.~\eqref{tau-cosm},
at the cosmological time $\tau_{cosm} = \tau_{BH}$ the cosmological plasma temperature was: 
\be
T_{cosm} (\tau_{BH})\approx 1.8\,{\rm eV} .
\label{T-cosm-tau-BH}
\ee
{It is known~\cite{H-rec} that the hydrogen recombination  temperature is $T_{rec} \approx 3100K \approx 0.27$~eV.
Correspondingly the red-shift at recombination is 
$z_{rec} = 1100$ and  the universe age at the recombination was $t_U^{(rec)} \approx 1.2\times 10^{13} $ sec. }
The recombination temperature  is 
 much lower than the hydrogen binding energy, $E_{bind} = 13.6$ eV, due to a large ratio 
of photon-to-baryon number densities, $n_\gamma /n_B = 1.6\times 10^{9}$, 
so there is always  a photon with energy larger than binding energy even at such low temperature of recombination.
Thus the considered black holes evaporated, while the  cosmic plasma was still ionized. 

We assume that the energy density of black holes is smaller than the total cosmological density at the cosmological
time equal to the BH life time and at $T_{cosm} = 1.8$ eV,
namely: 
\be 
M_{BH} n_{BH} < \frac{\pi^2 g_*}{30 }  T^4_{cosm}. 
\label{rho-BH-versus-rho-cosm}
\ee
It is realized if: 
\be
n_{BH} < \frac{\pi^2 g_*}{30}\,
\frac{(1.8\,{\rm eV})^4 }{10^{10} {\rm g}M_{10}} = 
\frac{\pi^2 g_*}{30M_{10}}\,\frac{1.8^4 \,10^{-36}\,{\rm GeV}^4} {5.62\cdot 10^{33} \,{\rm GeV} } 
\approx\frac{10^{-26}}{M_{10} {\rm cm}^{3}},
\label{rho-cosm-BH}
\ee
where we used the relations: $1\,{\rm g} = 5.62\cdot 10^{23} $ GeV and $ 1$ GeV $=5.06 \cdot 10^{13} $ cm$^{-1}$.

The temperature of cosmic plasma during the main part of evaporation period
was quite low, so the produced baryons could not destroy existing primordial deuterium, helium, and  lithium nuclei, since their binding 
energies are  several MeV, while the created anti-baryons could be dangerous and might destroy $^4$He and deuterium. 
However, below we show, that the produced anti-baryons efficiently annihilate with baryons of cosmological background.

The cross-section of helium  fission by anti-baryons was estimated in several 
works, see e.g. Refs.~\cite{sig1, sig2, sig3}.  The results are typically of the order of 
\be
\sigma_{fis} \approx
100\,{ fm}^2 \approx 10^{-24}  cm^2.   
\label{sigma-fiss}
\ee
The  fluxes of baryons and  anti-baryons, created by a single PBH, are the same and equal to 
$J_B =0.1 g_* f_B T^3_{BH} $, where $T_{BH} =10^3 \,{\rm GeV}/M_{10}$, see Eq.~(\ref{TBH}) and
$ f_B < 1$ is the relative 
fraction of (anti)baryons with respect to the total radiation emitted by the BH.  

The total rate of (anti)baryons production, $\dot n_B$, emitted  from the sphere of the area  $A = \pi r_g^2 $
($r_g = 2M_{BH} /M_{Pl}^2 \approx 1.5\cdot 10^{-18} M_{10} $ cm)
per unit time by black holes, with the number density $n_{BH}$,
is determined by the equation:
\be 
\dot n_B  + 3H n_B \approx \dot n_B = J_B A\, n_{BH} =\left[ 0.1 g_* f_B\cdot (10^3 \,{\rm GeV}/M_{10})^3 \cdot (\pi r_g^2)\right]\,\cdot n_{BH},
\label{dot-nB}
\ee 
where the factor in square brackets is the rate of baryon or antibarion production 
 by a single black hole. Note, that  the number densities of baryons and antibarions are initially the same.
We have neglected in Eq.~\eqref{dot-nB} the Hubble friction term because the cosmological expansion is slow in comparison with 
the characteristic time of the process.

Inserting numerical values into equation above we obtain:
\be
\dot n_B= (0.225\pi  g_* f_B) M_{10}^{-1} \left({10^{-27}} GeV^3 \cdot cm^2\right) n_{BH} \approx
(1.8 g_* f_B) M_{10}^{-1} n_{BH}\,\, GeV.  
\label{dot-nB-num}
\ee
{since $T_{BH}^3 \approx 10^9 \,({\rm GeV} /M_{10})^3$, see Eq. (\ref{TBH}).
Using the relation $1 $ GeV = $1.52 \cdot 10^{24} ${ sec}$^{-1}$},  we rewrite the above equation in seconds: 
\be
\dot n_B = (2.75 g_* f_B) M_{10}^{-1} 10^{24} n_{BH}\,\, { \rm{ sec}{^{-1}}}.
\label{dot-nB-sec}
\ee

Accordingly the number density of the (anti)baryons produced via the complete black hole evaporation is:
\be 
n_B = n_{\bar B}\approx \tau_{BH} \dot n_{BH} 
= \left( 4.4  f_B g_*\right)  M_{10}^2 \times 10^{26} n_{BH} ,
\label{n-B}
\ee
where the  value $\tau_{BH} = 160M_{10}^3$ sec is used, see Eq.~(\ref{tau-10}).

The rate of  $^4$He fission by anti-baryons is governed by the  equation:
\be
{\dot n(^4 He) = - \sigma_{fis} v^{\bar p}_{He} n_B n(^4 He)} ,
\label{Gamma-fiz} 
\ee
{where $v^{\bar p}_{He}$ is the relative velocity of antiprotons and helium nuclei, that is essentially thermally average proton velocity.
A reasonable guess for the fission cross-section is the size of the $^4He$ nuclei squared:
$l{({^4He})} = 1.68\times 10^{-13}$ cm~\cite{He-4-size}, i.e.  {$ \sigma_{fis} =2.8\times 10^{-26}$} cm$^2$.}

The temperature of the cosmic plasma at the moment of BH evaporation is given by Eq.~(\ref{T-cosm-tau-BH}):
$T_{cosm} (\tau_{BH})\approx 1.8$ eV and the cosmological time, Eq. (\ref{tau-cosm}),
at that period is:
\be
\tau_{cosm} = 0.5 \,{\rm sec} \left({\rm MeV}/{T_{cosm}}\right)^2  = 1.54 \cdot 10^{11}\, {\rm {sec}}.
\label{tau-cosm-number}
\ee

The relative number of destroyed helium-4 nuclei by anti-baryons during the cosmological time, 
can be estimated as:
{\be
r_{He} \equiv \frac{\Delta n(^4He) }{n(^4He)} = \sigma_{fis} v^{\bar p}_{He} n_B \tau_{cosm}\, 
\approx 6 \times 10^{22} \,(f_B g_*) M_{10}^2  n_{BH}\,  v^{\bar p}_{He} \,{\rm cm}^3. 
\label{delta-He-1}
\ee }

\begin{figure}[htbp]
		\begin{center}
		\includegraphics[scale=0.47]{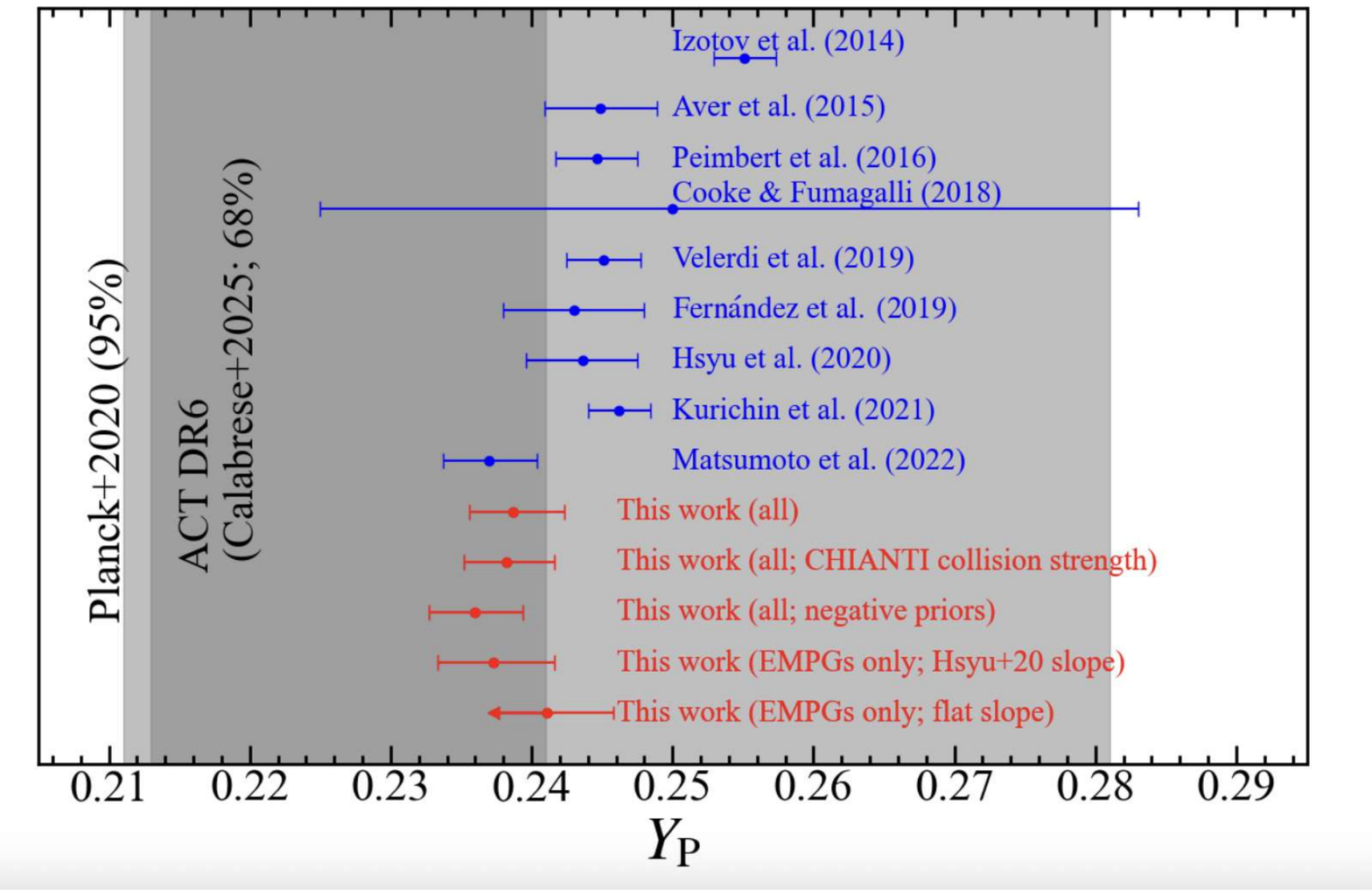} 
       \end{center}
	\caption{ {Primordial helium abundance taken from Ref.~\cite{Yanagisawa:2025mgx}.}
	}	
		\label{f-empr-2025}
\end{figure}

The results of the latest measurements of the primordial $^4$He abundace, $Y_P$, are presented in Ref.~\cite{Yanagisawa:2025mgx}, see Fig.~\ref{f-empr-2025} borrowed 
from the quoted paper. As is stated in this work the $Y_P$-value is lower than most of the previous estimates by one standard deviation. 

 In view of the presented data it is reasonable to assume that the uncertainty in determination of $^4He$ density is about 10\% , i.e. $r_{He} < 0.1$. Thus we obtain: 
\be
n_{BH} \lesssim \frac{2\cdot 10^{-24}\,{\rm cm}^{-3}}{(f_B g_*) \, v^{\bar p}_{He}\, M_{10}^2}.   
\label{n-BH-lim}
\ee

Consequently using this equation and Eq.~(\ref{n-B}) 
we find that the number density of excessive baryons 
is allowed to { be as large as:}
 {\be 
n_B = 4.4 \times 10^{26} \left(   f_B g_*\right)  M_{10}^2  n_{BH}\, { \lesssim }\,
{4.4 \times 10^{26} \left(   f_B g_*\right)  M_{10}^2 \frac{{ 2\cdot 10^{-24}}\,{\rm cm}^{-3}}{(f_B g_*)  \,  v^{\bar p}_{He}\, M_{10}^2}\,
{ \lesssim }\,  \frac{{10^3}}{ v^{\bar p}_{He}}\, {\rm cm}^{-3}}. 
\label{n-B-1}
\ee }

The cross-section of deuterium destruction is approximately the same as given by Eq.~\eqref{sigma-fiss}, so we expect 
similar variation of relative deuterium number density as presented in Eq.~\eqref{delta-He-1}:
\be
r_{D} \equiv \frac{\Delta n(D) }{n(D)} = \sigma_{fis} v^{\bar p}_{D} n_B \tau_{cosm}\, 
\approx 6 \times 10^{22} \,(f_B g_*) M_{10}^2  n_{BH}\,  v^{\bar p}_{D} \,{\rm cm}^3,
\label{delta-D-1}
\ee
where $v^{\bar p}_{D}$ is the relative velocity of antiproton and deuterium. 

Observational data on primordial deuterium abundances obtained by different groups are summarised in Table 4 of Ref.~\cite{Kislitsyn:2024jvk}. 
One can see that the relative variation of deuterium number densities exceeds 10\%. Therefore the account of the interaction of antibaryons with deuterium 
does not change restriction \eqref{n-BH-lim} on the black hole number density derived from the observations of $^4$He. 

\begin{figure}[htbp]
		\vspace{-2cm}
		\begin{center}
		\includegraphics[scale=0.5]{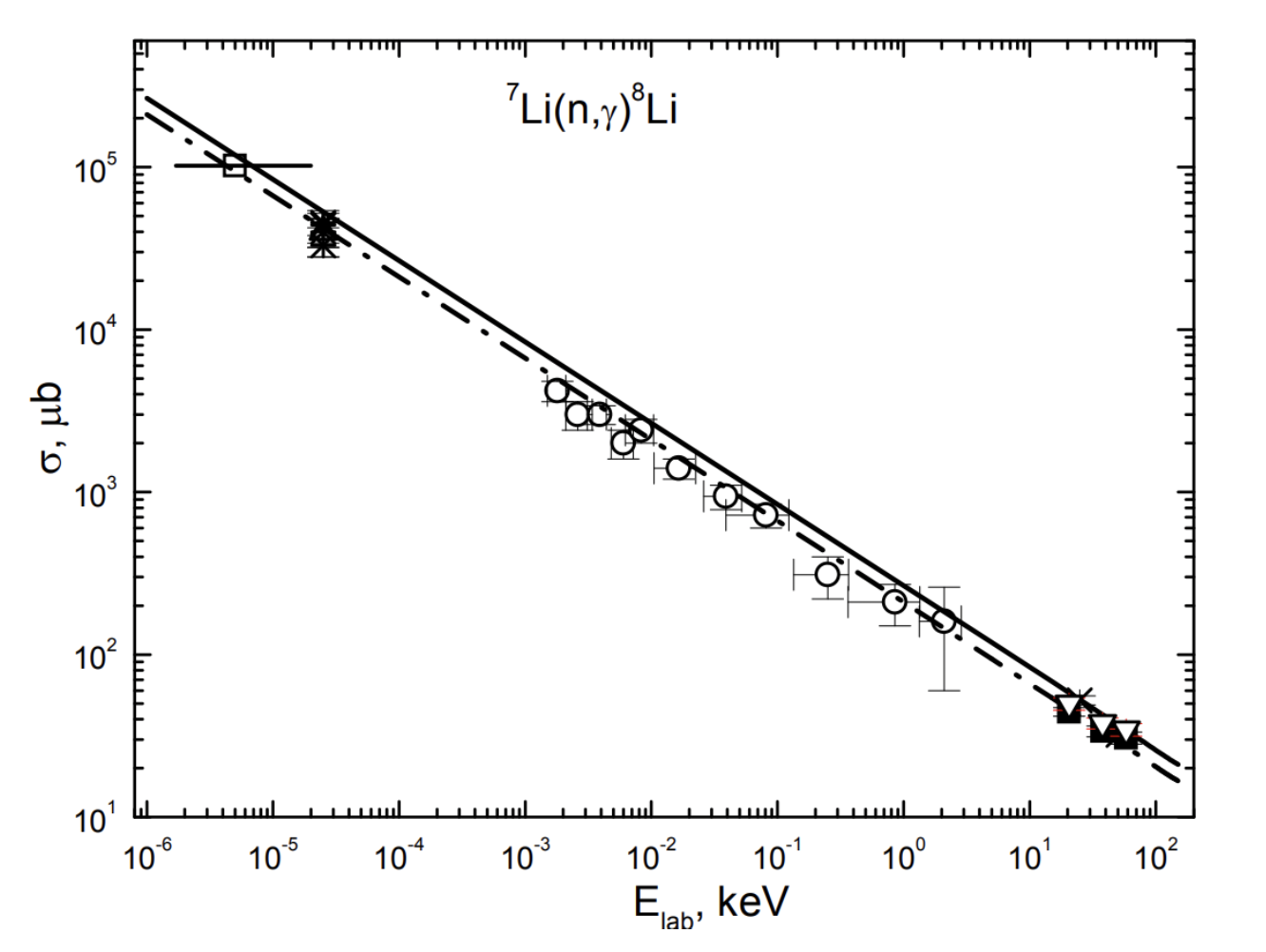} 
       \end{center}
	\caption{ { {Cross section { $^7$Li$+n \rar \gamma+^8$Li}, taken from Ref.~\cite{nLi7-gammaLi8}}.}
	}	
		\label{f-li-n}
\end{figure}

{We need to check now if this limited density of baryons 
originating from black holes is sufficient to destroy $^7$Li down to the observed value, see 
Fig.~{\ref{f-fig1}. The essential processes are $^7$Li $+n \rar  ^8$Li $+ \gamma$ with the cross-section 
{ $\sigma({^7Li}) \approx 2\cdot 10^{-25} {\rm cm}^2$,}  
see Fig.~\ref{f-li-n} borrowed from Ref.~\cite{nLi7-gammaLi8}. The solid curve is a theoretical estimate
of the authors and experimental points are from papers~\cite{Li-exp-1, Li-exp-2,Li-exp-3,Li-exp-4}.
Subsequently the produced  { $^8$Li} quickly decays into two $^4$He nuclei~\cite{Li-prob-rev}.}}


 The number density of $^7$Li evolves according to the equation:
\be 
\dot n_{Li}(t) = -  v_{Li}^n\sigma(^7Li)   n_{Li}(t) n_B, 
\label{dot-n-Li}
\ee
{where $n_{Li}(t)$ is the number density of $^7$Li and $ v_{Li}^n$ is the relative centre of mass velocity of the neutron and lithium nuclei
which is close to the thermal neutron velocity.}  

Eq.~\eqref{dot-n-Li} can be integrated leading to the result:
\be 
\frac{ n_{Li}(t)}{n_{Li}(t_{in})} = \exp \left( - \int _{t_{in}}^t\, dt' v_{Li}^n\sigma(^7Li)\,n_B \right).
\label{Li-int}
\ee

According to equation (\ref{Li-int})  the fraction of the destroyed $^7$Li during cosmological time 
(\ref{tau-cosm-number}) is  equal:
\be 
{r_{Li}} =
\frac{ n_{Li}(\tau_{cosm})}{n_{Li}(t_{in})} = \exp \left( -
\frac{v_{Li}^n}{v^{\bar p}_{He}}\,
\sigma({^7Li})   n_B \tau_{cosm} \right) \equiv e^{-K}. 
\ee
Substituting numerical values: $\sigma({^7Li}) \approx 2\cdot 10^{-25} {\rm cm}^2$,  $\tau_{cosm} = 1.54 \cdot 10^{11}$ sec, and $n_B$ from 
Eq.~\eqref{n-B-1}, we find the upper limit on $K$:
\be
K = 2\cdot 10^{ -25} cm^2 \times 10^3\, {\rm cm}^{-3}  \times 1.54 \cdot 10^{11} \times 3 \cdot 10^{10} \text{cm} \times \frac{v_{Li}^n}{v^{\bar p}_{He}}
\approx \frac{v_{Li}^n}{v^{\bar p}_{He}}.
\label{Delta-1}
\ee 
It is natural to expect that $  v_{Li}^n/v^{\bar p}_{He} \approx 1$. Therefore the $^7$Li abundance is supressed  by the factor $1/e \approx 0.37$. 

We see that with a reasonable choice of the parameters the density
of $^7$Li can be reduced to the observed value.

\section{ {Elimination of the evaporated antibaryons}}

In the previous section we assumed that the number density of antibaryons, which destroy  primordial nuclei, 
 is determined solely by the efficiency of the black hole evaporation. However, the evaporated antibaryons could annihilate with 
cosmic background baryons. If the annihilation is effective enough, the density of antibaryons could be strongly reduced and the derived above 
bounds on the BH density are lifted.    

Let us check that the number density of the cosmological background baryons, $n_B^{cosm}$, is larger than the number density of the evaporated 
antibaryons. The former is expressed through the number density of cosmic photons as: 
\be
n_B^{cosm} = \eta \, n_{\gamma} = 6 \cdot 10^{-10} \times 0.24\, T_{cosm}^3 = 1.44 \cdot 10^{-10}\, T_{cosm}^3. 
\label{n-B-cosm}
\ee

The  rate of annihilation of evaporated antibaryons with cosmological baryons is determined by the equation:
\be
\dot n_{\bar B} = -\sigma_{ann} v_{B \bar B} \,n_{\bar B} \, n_B^{cosm},
\ee
where $\sigma_{ann}$ is the $B \bar B$-annihilation cross-section, $v_{B \bar B}$ is the relative 
velocity of baryons and antibaryons.
In the limit of vanishing velocity  the product $\sigma_{ann} v_{B \bar B}$ tends to a constant value. We take it equal to 
the Compton wavelength of pion squared, i.e.  
$\sigma_{ann} v_{B \bar B} \approx 2 \cdot 10^{-26} $cm$^2$.  

This equation can be integrated as:
\be
\frac{n_{ \bar B} (t)}{n_{\bar B} (t_{in}) }  = \exp\left(-  \int^t_{t_{in}}  dt' \sigma_{ann} v_{B \bar B} \, n_B^{cosm}\right).
\label{n-B-of-t}
\ee
The asymptotic value of the antibaryon supression factor achieved during cosmological time is equal to:
\be 
{r_{\bar B}} = \frac{n_{ \bar B} (\tau_{cosm})}{n_{\bar B} (t_{in}) } = 
\exp\left(- \sigma_{ann} v_{B \bar B} \, n_B^{cosm}\,\tau_{cosm}\right) \approx e^{-10}.
\label{r-bar-b}
\ee 
One can see that the antibaryons emitted by evaporating black hole practically disappear and do not have destructive impact on the 
primordial nuclei abundances.

\section{Cool black holes \label{ss-cool-BH}}

In this section we consider the case of cooler black holes with e.g. $T_{BH} \leq$100 MeV or below, when
baryons are not  created. The only stable particles that can be produced by such black holes are electrons, positrons (in equal amount, if BH is 
electrically neutral), photons, and neutrinos. As follows from Eq.~(\ref{TBH}), in this case the black hole mass 
should be:
$ 
M_{BH} 
\approx 10^{14} {\rm g} \, (100 MeV /T_{BH}) $   
and its life-time would be longer than
\be
\tau_{BH} = 1.6 \cdot 10^{14}\,{ {\rm sec}}  \left( \frac{100 \, MeV}{T_{BH}}\right)^2 \gtrsim 1.6  \cdot 10^{14}\, { {\rm sec}}. 
\label{tau-BH-100}
\ee

The photon energy immediately after their production from the BH evaporation were close to the BH temperature. However,
the photons could be thermalised due to their scattering on the cold background electrons.
The temperature of the cosmological plasma at the cosmological time equal to the BH age (\ref{tau-BH-100}) 
would be below the value:
\be
T_{cosm}(\tau_{BH})
 \lesssim \left( \frac{0.5}{1.6\cdot 10^{14}} \right)^{1/2} \, \left(\frac{T_{BH} }{100\, MeV}\right)\, MeV =
0.06\, {\rm eV} \left( \frac{T_{BH} }{100\, MeV} \right).
\label{T-cosm-100}
\ee
This temperature is noticeably smaller than the hydrogen recombination temperature, 
$T_{rec} \approx 0.26 $ eV~\cite{H-rec}
 that took place at the redshift $z_{rec} = 1100$.
So the energetic gamma quanta interacting with neutral atoms would lead to plasma reionisation at redshifts 
$z_{reion} \lesssim 20$, adding possibly noticeable contribution to the canonical sources of reionisation, see e.g. Ref.~\cite{reion}.
Another potentially observable effect could be a distortion of  the CMB spectrum at low frequencies.

Photon thermalisation by scattering on electrons would diminish their energy below $^4$He binding energy so they would not destroy
helium-4 nuclei  but they could diminish the number densities of lithium and beryllium down to the observed values.

\section{ Summary} 

It is shown that a population of evaporating primordial black holes may successfully solve the problem of the too low observed value of 
of lithium number density at the canonical magnitude of the cosmological baryon density. There are two possibilities of "hot" and "cool"
black holes. In the first case excessive  lithium is killed by nucleons, 
through the reactions  $^7$Li$+n \rar ^8$Li $+ \gamma$  or $^7$Li $+ p \rar ^8$Be$ + \gamma$.
Neutrons produced by black holes quickly loose their energy interacting with the primeval plasma.
Subsequently both $^8$Li and $^8$Be quickly decay into a pair $^4$He nuclei.

In the second case the depletion of $^7$Li  could be driven by photons.

The considered mechanisms may lead to distortion on the CMB angular fluctuations at low frequencies and to
make contribution to reionization of the universe.


\section*{Acknowledgments}
This work  was supported by the RSF grant 23-42-00066.

\end{document}